\documentstyle[12pt]{article}
\begin{document}           
\title{Singularities and asymptotic behavior of the Tolman-Bondi model
\thanks{
Report-no: IRB-StP-GR-050797
}}
\author{
Alexander Gromov \thanks{
St. Petersburg State Technical University,
Faculty of Technical Cybernetics, Dept. of Computer Science,
29, Polytechnicheskaya str. St.-Petersburg, 195251, Russia
{\it and}
Istituto per la Ricerca di Base,
Castello Principe Pignatelli del Comune di Monteroduni,
I-86075 Monteroduni(IS), Molise, Italia.
E-mail: gromov@natus.stud.pu.ru
}}
\date{}
\maketitle
\begin{abstract}
The Bondi formula for calculation of the invariant mass in the
Tolman- Bondi (TB) model is interprated as a transformation rule on the set
of co-moving coordinates.
The general procedure by which the three arbitrary functions
of the TB model are determined explicitly is presented.
The properties of the TB model, produced by the transformation rule are
studied.

Two applications
are studied:
for the falling TB flat model the equation of motion of two singularities
hypersurfaces
are obtained;
for the expanding TB flat model the dependence of size of area with
friedmann-like solution on
initial conditions is studied in the limit $t \to +\infty$.
\\ PACS number(s): 95.30.Sf, 98.65.-r, 98.80.-k
\\ key words: cosmology:theory --- gravitation ---large- scale structure of
universe --- relativity
\end{abstract}

\section{Introduction} \label{1}

The Tolman-Bondi (TB) model (Tolman, 1934; Bondi, 1947) describes
a spherical symmetry self consistency dust motion in a co-moving
coordinates. The Einstein's equations are reduced to the definition of the
coefficient $g_{1,1}$ of the metrical tensor, the second order
partial differential equation of motion for the function $R(r,t)$ (the
analog of Euler coordinate) and the algebraic density law.

The Tolman-Bondi (TB) model depends on three undetermined functions: $f(r)$,
$F(r)$ and ${\rm F}(r)$ of a co-moving coordinate $r$.
One of them, $f(r)$, defines the $g_{1,1}$
component of the metric and two others are obtained in the process of
integration of the equation of motion.

The TB model is defined up to a transformation $\Phi: r \to \tilde r$.
But the TB model is also a hydrodynamical problem which requires only two
undetermined function to start with: the initial density and velocity
distribution.
In the set of publications (Ruban and Chernin, 1969; Bonnor, 1972; Bonnor,
1974; Silk 1977; Gurevich and Chernin, 1979) the number of undetermined
functions decreases by the use of the different transformation of co-moving
coordinates (see table 1). The common property of these transformations is
the definition of new coordinates through not more that the only one of
three undetermined functions of the TB model.

The main idea of this paper is the interpretation of
the the invariant mass formula, calculated by Bondi (1947), as the
transformation rule from an arbitrary co-moving coordinates $r$ to the
invariant mass coordinate $M_i$. The $M_i$ coordinate is singled out by its
physical sense. The coordinate $M_i$ will be called Bondi coordinate and
the definition of invariant mass in the TB model will be called Bondi
transformation.

Unlike other transformations, the Bondi transformation is defined
on the couple of undetermined functions $(f,F)$, so that the dependence $F
= F(f)$ is provided. It is the new way to decrease the number of
undetermined functions and the way is unique for all meanings of $f$.

The Bondi transformation produces also: the dependence of initial condition
$R_0(r)$ for the function $R(r,t)$ on the function $f$ and initial
density distribution $\rho_0(r)$: $R_0 = R_0(f,\rho_0(r))$;
the integral equation for the function $f$ with coefficients $\rho_0(r)$
and initial velocity profile $\dot R_0(r)$;
the factorisation on the set of couples of functions $(f,F)$;
the factorisation on the set of couples of initial conditions $(\dot
R_0(r),\rho_0(r))$.
\begin{table}
\begin{center}
\begin{tabular}{|c|c|}   \hline
author       & $\Phi: r \to \mbox{new coordinate}$ \\ \hline
Ruban and Chernin,1969 & $r \to M(r)$ \\
Bonnor, 1972 & $r \to \alpha \, R_0(r)$ \\
Bonnor, 1972 & $r \to \alpha \,\left(1 + \alpha \,r^3\right)^{-1/6}$ \\
Bonnor, 1974 & $r \to \alpha\,F^{1/3}(r)$ \\
Silk, 1977 &     $r \to F(r)/(2\,E(r))$\\
Gurevich and Chernin 1979 &$r \to F(r)/4$ \\
\hline
\end{tabular}
\end{center}
\caption{
Examples of co-moving coordinates.
$M$ is the mass(in the Newtonian theory); $r$ is the arbitrary co-moving
coordinate; $R_0(r)$ is the analog of Euler radial coordinate;
$\alpha$ is the constant; $E(r)$ is
the full specific energy.
}
\end{table}

The obtained results are applied to study several properties of the flat TB
model.
For the falling flat models the equations of the singularity
hypersurfaces are obtained. It is shown how the gradient of the initial
density distribution produces the split of one singularity of FRW model to
two singularities of the TB model.
For the expanding TB models the dependence of a character of
friedmannisation on the initial density distribution is studied.

First the statement about friedmannisation of the expanding TB models
has been presented in the article Collins and Hawking (1973). They
have studied the
homogeneous cosmological
models but they noted that although they work '... does not prove that
there  is no open set of inhomogeneous initial data which gives rise to
models that approach homogeneity and isotropy ... it makes it appear very
improbable since one would expect inhomogeneities to produce anisotropy
rather that isotropy'.
Bonnor (1974) has presented the counterexample and sowed that
'... an expanding Tolman model with $f^2 = 1$ and everywhere
non-vanishing density necessarily avolves to the homogeneous Einstein- de
Sitter model, whatever its initial condition.'
This paper studies the same problem from the viewpoint of the results
obtained on the ground of the Bondi transformation.

The short review of the TB model is presented in the section \ref{2}.
The interpretation of the Bondi formula for calculation the invariant mass
as a transformation rule and the definition of the functions $F$ and ${\bf
F}$ on the base of the transformation rpesents in the section \ref{3}.
In the section \ref{4} one initial condition for the equation of motion
and the equation for function $f$ are obtained; the partial case: FRW model
is discussed. The sections \ref{5} and \ref{6} are dedicated to the
application of the obtained results to the falling and expanding flat TB
models. The article is ended by the summary.

\section{The TB Model} \label{TB model} \label{2}

The Tolman-Bondi model (TB) is the simplest spherical symmetry
inhomogeneous model in general relativity (Tolman, 1934; Bondi, 1947).
TB model describes pressureless dust motion
in a co-moving coordinates in the metric ($c = 1, G = 1$)
\begin{equation}
{\rm d}s^2(r,t) = {\rm d}t^2 - \frac{R^{\prime\, 2}(r,t)}{f^2(r)}\,{\rm
d}r^2 - R^{2}(r,t)\,{\rm d} \Omega^2,
\quad {}^{\prime} = \frac{\partial}{\partial r},
\label{metric}
\end{equation}
where $R(r,t)$ is an analog of Euler coordinate,
$f(r)$ is the undetermined function of the model, obtained by the
integration of (1,4) component of the Einstein's equations. Others
equations give the equation of motion
\begin{equation}
2\,\ddot R(r,t)R(r,t) + \dot R^2(r,t)
+1 - f^2(r) = 0
\label{eq-m}
\end{equation}
and the density law:
\begin{equation}
8 \pi \rho(r,t) = \frac{{\rm d} F(r)}{{\rm d}
r}\,\frac{1}{2\,R^2(r,t)\,\displaystyle\frac{\partial R(r,t)}{\partial r}}.
\label{T:15:1}
\end{equation}
The equation of motion is solved together with initial conditions:
\begin{equation}
\left.R(r,t)\right|_{t = 0} = R_0(r),
\label{Bb-1}
\end{equation}
\begin{equation}
\left.\dot R(r,t)\right|_{t = 0} = \dot R_0(r),
\label{Bb-2}
\end{equation}
here $R_0(r)$ and $\dot R_0(r)$ are specified functions.
The first integral of the equation of motion is
\begin{equation}
\frac{1}{2}\dot R^2(r,t) =
\frac{f^2(r) - 1}{2} +
\frac{F(r)}{4\,R(r,t)},
\label{TB-klass}
\end{equation}
where $F(r)$ is the second undetermined function.
Bondi got the interpretation of two functions $f$ and $F$ on the ground
of the comparison of the equation (\ref{TB-klass}) with Newtonian theory:
$\frac{f^2(r) -1}{2}$ may be interpreted as an analog of a full specific
energy $E_0(r)$ and
$F(r)/4$ may be interpreted as an analog of an effective gravitating mass.

The integral of the equation (\ref{TB-klass}) has the form
\begin{equation}
\pm t + {\bf F}(r) =
\int\frac{{\rm d} \tilde R}{
\sqrt{
f^2(r) - 1 + \frac{F(r)}{2\,\tilde R}}},
\label{T:15:11}
\end{equation}
where ${\bf F}(r)$ is the third undetermined function, the sign $'+'$
corresponds to an expanding solution and the sign $'-'$ corresponds to the
falling one.

The functions $f(r)$, $F(r)$ and ${\bf F}(r)$ are three undetermined
functions of the TB model.

\section{The Definition of the Functions $F$ and ${\bf F}$} \label{3}

Ruban and Chernin (1969) have studied the Newtonian analog of
the TB model in the coordinatisation, defined by the equality
\begin{equation}
r = M.
\label{Ruban&AD}
\end{equation}
It was shown that the use of this equality decreases a number of
undetermined functions by one unit.

Bondi showed that in the Newtonian theory invariant mass and gravitating
mass coincide, but in the TB model it is not true (Bondi, 1947).
The using of the effective gravitating mass coordinatisation given by $F/4$
has been represented in the set of articles (for example, Gurevich and
Chernin (1979), Silk (1977) and it is also the way to decrease the number
of indeterminated functions from 3 to 2.

In this paper the invariant mass $M_i$, calculated by Bondi
(1947):
$$M_i(r) = 4\,\pi\,\int\limits_{0}^{r}\rho(r,t)\,\sqrt{-g(r,t)}\,{\rm d}r =$$
\begin{equation}
4\,\pi\,\int\limits_{0}^{r}\rho(r,t)
\frac{R^2(r,t)}{f(r)}\frac{\partial R(r,t)}{\partial r}
\,{\rm d}r =
\frac{1}{4}\int\limits_{0}^{r}\frac{F^{\prime}(r)}{f(r)}{\rm d} r,
\quad F^{\prime}(r) = \frac{{\rm d} F(r)}{{\rm d} r},
\label{Bondi}
\end{equation}
is used as independent co-moving variable.
We will name $M_i$ Bondi coordinatisation.
The fact that $M_i$ is independent from time means that the particle
layers do not intersect.
The definition of invariant mass has a form of a transformation rule from
$r$ to $M_i$. This transformation will be called Bondi transformation.
The identical Bondi transformation
\begin{equation}
r = M_i
\label{Ruban&AD-i}
\end{equation}
produces the relation between two functions $f$ and $F$:
\begin{equation}
F(M_i) = 4\,\int\limits_{0}^{M_i}f(M_i)\,{\rm d}M_i.
\label{F def}
\end{equation}
>From this point in this paper
we will use the Bondi coordinatisation.
>From (\ref{Bondi}) and (\ref{F def}) it follows that the coordinatisation
$M_i$ is singled out by the fact that the invariant
density is constant:
\begin{equation}
\rho(M_i,t)\,\sqrt{-g(M_i,t)} = \frac{1}{4\,\pi}.
\label{i-d-R-Ch}
\end{equation}

Now let us study the case of two coordinatisations:
an undetermined coordinatisation $r$ and $M_i$.
Suppose the functions $f(r)$ and $F(r)$ are specified.
The Bondi transformation is the example of the continuous transformation up
to which the coordinatisation is defined.
So, we have three functions $f(r)$, $F(r)$ and $M_i(r)$, any two of which
define the third function.

Tolman (1934) also uses the third undetermined function
${\bf F}$ to solve the equation (\ref{TB-klass}).
The function ${\bf F}(M_i)$ in the coordinatisation $M_i$  has the form
\begin{equation}
{\bf F}(M_i) =
\int\frac{{\rm d} \tilde R}{
\sqrt{
f^2(M_i) - 1 + \frac{2}{\tilde R}\,\int\limits_0^{M_i} f(M_i)\,{\rm d}M_i}}.
\label{T:12}
\end{equation}
We can see now that two functions: $F$ and ${\bf F}$ depends on the
function $f$.

\section{The Initial Conditions for the TB Model} \label{i-c} \label{4}

Let us suppose that an initial profile of density is given in the Bondi
coordinatisation. From (\ref{Bondi}) it follows
that
\begin{equation}
R(M_i,0) =
R_{0}(M_i) = \left[
\frac{3}{4\,\pi}\,\int\limits_{0}^{M_i}
\frac{f(\tilde M_i)}{\rho(\tilde M_i,0)}\,{\rm d} \tilde M_i
\right]^{1/3}.
\label{Mm}
\end{equation}
>From (\ref{Mm}) and (\ref{F def}) it follows that $f \ge 0$ and $F \ge 0$.
Using (\ref{F def}) and (\ref{TB-klass}) in the initial moment of time
$t = 0$, together with the initial condition (\ref{Mm}), we obtain the
following equation for the function $f(M_i)$:
\begin{equation}
\frac{1}{2}\,\left(\frac{3}{4\,\pi}\right)^{1/3}\,
\left(\int\limits_{0}^{M_i}
\frac{f(M_i)\,{\rm d}M_i}{\rho_0(M_i)}\right)^{1/3}\,
\left[\dot R^2_0(M_i) - f^2(M_i) + 1\right]  =
\int\limits_{0}^{M_i}
 f(M_i){\rm d}M_i.
\label{f def}
\end{equation}
The solution of this equation has the form
\begin{equation}
f(M_i) = f \left(\rho_0(M_i), \dot R_0(M_i)\right).
\label{f sol}
\end{equation}
>From (\ref{Mm}) it follows that $f(M_i) \ge 0$ for every initial density
distribution $\rho_0(M_i)$ , and from (\ref{F def}) it follows that $F \ge
0$.
This nonequality implies $\dot R^2_0(M_i) - f^2(M_i) + 1 \ge 0$.

The equation should be used as follows: we start from the initial conditions
$\rho_0(M_i)$ and $\dot R_0(M_i)$ and calculate the function $f(M_i)$.
Then we
fix the function $f(M_i)$ and (\ref{f def}) becomes the definition of
the set of initial conditions which do not change the function $f(M_i)$.

Substituting (\ref{F def}) and (\ref{Mm}) in (\ref{T:15:1}) we see that the
density law becomes the identity in the initial moment of time $t = 0$.
So, the initial conditions of the
TB model coincide with the initial conditions of the equation of
motion (\ref{eq-m}).

In case of FRW model the initial condition (\ref{Mm}) becomes
\begin{equation}
R_{FRW}(M_i,0) = \left[
\frac{3}{4\,\pi\,\rho_{FRW}(0)}
\int\limits_{0}^{M_i}
f(\tilde M_i)\,{\rm d} \tilde M_i
\right]^{1/3}.
\label{FRW-2}
\end{equation}
In case of $f = 1$ and $\rho(M_i,0) = \rho_0 = const$ (\ref{f def}) is
reduced to
\begin{equation}
3\,H^2 = 8\,\pi\rho_0,
\label{FRW-3}
\end{equation}
where
\begin{equation}
H = \left.\frac{\dot R_0(M_i)}{R_0(M_i)}\right|_{FRW} =
\sqrt{\frac{8\,\pi\,\rho_0}{3}} = const.
\label{FRW-4}
\end{equation}
On the other hand, from the assumption that
\begin{equation}
\dot R_0(M_i) = H\,R_0(M_i),
\label{FRW-5}
\end{equation}
where $H = const$, from (\ref{f def}) it follows that
\begin{equation}
\rho_0(M_i,0) = const.
\label{FRW-6}
\end{equation}

\section{Application for Falling Flat Solution} \label{functions} \label{5}

In this section we will study several properties of the falling flat ($f^2
= 1$) TB model.
The solution of the equation of the TB model with $\Lambda = 0$ has been
obtained by Bonnor (1972, 1974).
The flat Bonnor's solution is obtained by the substitution of $f^2 = 1$
into (\ref{TB-klass})  and the following equation is obtained:
\begin{eqnarray}
R(M_i,t\,)\left(\frac{\partial R(M_i,t)}{\partial t}\right)^2 =
2\,M_i.
\label{TB-2u}
\end{eqnarray}
This equation is studied together with the initial conditions (\ref{Mm}).
The solution is as follows:
\begin{eqnarray}
R^{3/2}(M_i,t) = R_0^{3/2}(M_i) - 3\,\sqrt{\frac{M_i}{2}}\,t.
\label{TB-3ugu}
\end{eqnarray}
Note here, that from (\ref{TB-3ugu}) and (\ref{Mm}) it follows that the
possibility of representation of the $R(M_i,t)$ as the multiplication of
the two functions, one of which depends only on $M_i$ and the second one,
which depends only on $t$,
is truly only for FRW model:
\begin{eqnarray}
R_{FRW}^{3/2}(M_i,t) = 3\,\sqrt{\frac{M_i}{2}}\,
\left(\frac{1}{\sqrt{6\,\pi\,\rho_0}}
- t \right).
\label{ab}
\end{eqnarray}

The density law of the flat TB model is obtained by the substitution of
(\ref{TB-3ugu}) into (\ref{T:15:1}) and has the form:
\begin{eqnarray}
4\,\pi\,\rho(M_i,t) = \frac{1}{
\left[
R^{3/2}_0(M_i) - \frac{3\, t}{\sqrt{2}}\,\sqrt{M_i}
\right]
\left[
\sqrt{R_0(M_i)}\,R^{\prime}_0(M_i) - \frac{t}{\sqrt{2\,M_i}}\,
\right]
}.
\label{den-flat}
\end{eqnarray}
The density has the singularity when one of the parts of the equation in
parenthesis is equal to zero.
Two singlarities are in accordance with two difference reasons: the falling
of the dust to the cenrte $R = 0$ is described by the equation
\begin{eqnarray}
R^{3/2}_0(M_i) - \frac{3\, t_1(M_i)}{\sqrt{2}}\,\sqrt{M_i} = 0;
\label{den-sing-1}
\end{eqnarray}
the breakdown of the condition of particle layers nonintersecting
equivalents to the equation
\begin{eqnarray}
\sqrt{R_0(M_i)}\,R^{\prime}_0(M_i) -
\frac{t_2(M_i)}{\sqrt{2M_i}} = 0.
\label{den-sing-2}
\end{eqnarray}
$t_1(M_i)$ and $t_2(M_i)$ are the hypersurfaces where
$\rho(M_i,t) = \infty$.

Substituting (\ref{Mm}) into (\ref{den-sing-1}) and (\ref{den-sing-2}) we
obtain the equations of the hypersurfaces:
\begin{eqnarray}
t_1(M_i) = \sqrt{
\frac{1}{6\,\pi\,M_i}\,\int\limits_0^{M_i}\frac{{\rm d}\tilde
M_i}{\rho_0(M_i)}
},
\label{surf-sing-1}
\end{eqnarray}
\begin{eqnarray}
t_2(M_i) = \frac{1}{\rho_0(M_i)}
\sqrt{
\frac{M_i}{6\,\pi\,\int\limits_0^{M_i}\frac{{\rm d}\tilde
M_i}{\rho_0(M_i)}}}.
\label{surf-sing-2}
\end{eqnarray}
>From (\ref{surf-sing-1}) and (\ref{surf-sing-2}) it follows
\begin{eqnarray}
t_1(M_i)\,t_2(M_i) = \frac{1}{6\,\pi\,\rho_0(M_i)}.
\label{t1 t2}
\end{eqnarray}
In the limit of $\rho(M_i,t) = \rho(t)$ we obtain
\begin{eqnarray}
t_1(M_i) = t_2(M_i) = \frac{1}{\sqrt{6\,\pi\,\rho(t)}}.
\label{limit}
\end{eqnarray}
This means that a gradient in the initial profile of the
density splits one singularity of the FRW model into two singularities in
the TB model.

\section{Application for Expanding Flat Solution} \label{6}

In this section we apply the results of sections (\ref{i-c}) and
(\ref{functions}) to the expanding TB flat model.

Bonnor (1974) showed that the set of flat TB models evolve
asymptotically towards a FRW model. This is meant to be a counterexample to
the statement of Collins and Hawking (1973) that
measure zero of all models can evolve towards FRW.

Using the initial conditions in the form (\ref{Mm}) and Bondi
coordinatisation $M_i$  we will study here the influence of different
initial conditions on the character of friedmannisation of the Tolman-Bondi
flat models.

\subsection{Bonnor's initial condition}

First we will analyze the initial conditions given by Bonnor
(1974).
Bonnor uses the flat solution of the TB model in the form:
\begin{eqnarray}
R(M_i,t) = \frac{1}{2}\,\left[9\,F(M_i)\right]^{1/3}\left[
t + \beta(M_i)
\right]^{2/3}.
\label{LTB-R15}
\end{eqnarray}
Using (\ref{F def}) together with $f^2 = 1$ and comparing (\ref{TB-3ugu})
and (\ref{LTB-R15}) we find out that
\begin{eqnarray}
R(M_i,t) = \frac{9^{1/3}}{2}\,\left(4\,M_i\right)^{1/3}
\left[t + \beta(M_i)
\right]^{2/3},
\label{LTB-R1}
\end{eqnarray}
where
\begin{eqnarray}
\beta(M_i) = \frac{1}{3}\,\sqrt{\frac{2\,R^3_0(M_i)}{M_i}}
\label{new-beta}
\end{eqnarray}
Using the initial conditions (\ref{Mm}) we obtain
\begin{eqnarray}
\beta(M_i) = \sqrt{\frac{1}{6\,\pi\,M_i}\,
\int\limits_{0}^{M_i}\frac{{\rm d}\,\tilde M_i}{\rho_0(\tilde M_i)}}.
\label{new-beta-1}
\end{eqnarray}

Bonnor (1974) studies one particular case of initial conditions
\begin{eqnarray}
\beta_0 = \beta(0) = 0.
\label{LTB-B}
\end{eqnarray}
It follows from (\ref{new-beta-1}) that
\begin{eqnarray}
\beta(0) = \frac{1}{\sqrt{6\,\pi\,\rho_0(0)}}
\label{LTB-Bu}
\end{eqnarray}
so, (\ref{LTB-B}) means
\begin{eqnarray}
\rho_0(0) = \infty.
\label{LTB-Bubu}
\end{eqnarray}
At the same time it follows from (\ref{LTB-R1}) and (\ref{LTB-B}) that
\begin{eqnarray}
R(M_i,0) = 0,
\label{LTB-ubu}
\end{eqnarray}
and there are no particles in the area $M_i > 0$.
It means that the Bonnor's initial condition (\ref{LTB-B}) produces the
TB model with delta-like distribution of dust:
\begin{eqnarray}
\rho_0(M_i) = \delta(M_i).
\label{i=ini}
\end{eqnarray}

Bonnor has used the equation (\ref{LTB-B}) to prove the statement:
'... an expanding Tolman model with $f^2 = 1$ and everywhere
non-vanishing density necessarily avolves to the homogeneous Einstein- de
Sitter model,
whatever its initial condition.' Using the Bonnor transformation, we saw
that this statement is true only for delta-like initial density
distribution.

\subsection{Non delta-like initial conditions}

We will study now how the non delta-like initial condition for density
\begin{equation}
\rho(M_i,0) = \rho_0(M_i)
\label{non-delta}
\end{equation}
change the result.
This initial condition produces new asymptotic properties of the TB model.

We will study the following class of initial conditions of the
density distribution: the density monotonically decreases with increase of
$M_i$ and
\begin{equation}
\lim_{R(M_i) \to +\infty} \rho_0(M_i) = 0
\label{prop-1}
\end{equation}
together with
\begin{equation}
\lim_{R(M_i) \to +\infty}
\frac{1}{M_i}\,\int\limits_{0}^{M_i} \frac{{\rm d}\tilde M_i}{\rho_0(\tilde M_i)}
 = +\infty.
\label{prop-2}
\end{equation}
The left side of the equation is the average of $1/\rho_0(M_i)$.
For this class of initial conditions
\begin{equation}
\lim_{R(M_i) \to +\infty} t_1(M_i) = +\infty
\label{prop-3}
\end{equation}
and
\begin{equation}
t_2(M_i) > t_1(M_i),
\label{prop-4}
\end{equation}
so
\begin{equation}
\lim_{R(M_i) \to +\infty} max\{t_2(M_i), t_1(M_i)\} = + \infty.
\label{prop-5}
\end{equation}

The inversion of the equations (\ref{surf-sing-1}) and (\ref{surf-sing-2})
has the form
\begin{equation}
M_1 = M_1(t) \quad \mbox{and} \quad M_2 = M_2(t),
\label{prop-6}
\end{equation}
and describes two surfaces in the Tolman space-time with the following properties:
the surface $M_1(t)$ separates the area $0 < M_i < M_1(t)$ where the initial
conditions are forgotten for the function \newline
$R(M_i,t > t_1(M_i))$ and the area
$M_1(t) < M_i$, where it is not possible to neglect the initial
conditions; similarly,
in the area $0 < M_i < M_2(t)$ the initial conditions are forgotten
for $R^{\prime}(M_i,t)$ and for time $t > t_2(M_i)$.
For the velocity ${\rm d}\,M_1/{\rm d}\,t$ we obtain:
\begin{equation}
\lim_{R(M_i) \to +\infty}\,\frac{{\rm d} M_1(t)}{{\rm d} t} = 0.
\label{velocity}
\end{equation}

The conditions (\ref{prop-5}) and (\ref{velocity})
show that for the chosen class of the initial
conditions both surfaces $M_1(t)$ and $M_2(t)$ never
arrive in the point
where $\rho_0(M_i) = 0$.

Let us calculate now the density in the two areas:
\begin{equation}
0 < M_{core} < min\{M_1(t),M_2(t)\}
\label{area: core}
\end{equation}
and
\begin{equation}
max\{M_1(t),M_2(t)\} < M_{shell} < +\infty.
\label{area: shell}
\end{equation}
The density law for the expanding model is:
\begin{eqnarray}
\rho(M_i,t) = \frac{1}{6\,\pi\,\left[t_1(M_i) + t\right]\,
\left[t_2(M_i) + t\right]},
\label{den-flat-expan}
\end{eqnarray}
where $t_1(M_i)$ and $t_2(M_i)$ are represented by (\ref{surf-sing-1}) and
(\ref{surf-sing-2}) respectiveles.
At the limit $t \to +\infty$ from (\ref{den-flat-expan}) it follows that
\begin{eqnarray}
\rho_{core}(M_i,t) \approx \frac{1}{6\,\pi\,t^2}
\label{den-flat-core}
\end{eqnarray}
and
\begin{eqnarray}
\rho_{shell}(M_i,t) \approx
\frac{1}{4\,\pi\, R^{2}_0(M_i)\,R^{\prime}_0(M_i)} = \rho_0(M_i).
\label{den-flat_shell}
\end{eqnarray}
The both initial conditions  for $R(M_i,t)$
and for $R^{\prime}(M_i,t)$ are forgotten simultaneously in the core
in the limit of
\begin{eqnarray}
t \gg max\{t_1(M_i),\,t_2(M_i) \}.
\label{core-time}
\end{eqnarray}

Because the both surfaces $M_1(t)$ and $M_2(t)$ never arrive in to the
point with $\rho_0(M_i) = 0$
the area of friedmannisation of the TB flat expanding model is also
limited by the nonequality
\begin{eqnarray}
M_{core} \ll min \{ M_1(t), M_2(t) \}.
\label{lim}
\end{eqnarray}

\section{Summary}

The use of the definition of the invariant mass allows to define the
transformation rule on the set of co-moving coordinates. The identical
transformation is studied in this paper.

The indentical transformation produces the initial conditions for the
function $R_0(M_i)$ is given by (\ref{Mm}) and the definition of the
function $f(M_i)$ is given by (\ref{f def}).
The equation has the solution in the form $f = f(\rho_0(M_i),\dot
R_0(M_i))$, so the function $f$ defines the TB model.
For fixed TB model, i.e. for fixed
function $f$, the equation (\ref{f def}) becomes the definition of
the set of equivalent couples of initial conditions $(\rho_0(M_i),\dot
R_0(M_i))$.

The flat models are studied together with initial conditions (\ref{Mm}).
For the falling flat models the equations of the singularity
hypersurfaces are studied. It is shown how the gradient of the initial
density distribution produces the split of one singularity of FRW model to
two singularities of the TB model. For the expanding flat model it shown
that only delta-like initial density distribution produces the FRW model in
the limit of $t \to +\infty$. For the chosen class of initial conditions
non delta-like initial density
distribution go to the FRW-like core and TB-like shell in that limit.

The obtained solutions for the falling and expanding flat models are
stronly depends on the initial conditions but independs from
the value of
function $f$. It means that every expanding (i.e. non flats) TB models also
have the friedmann-like core and nonfriedmann-like shell in the limit $t
\to +\infty$.
The difference is in the structure of the TB-like shell.
The motion of the singularitiy hypersurfaces (for falling TB
models) has a different
character for different initial density profiles and can brake the toplogy
of the area of definition of the TB model.

\section{Acknowledgements}

I'm grateful for encouragement and discussion
to Prof. Arthur Chernin,  Prof. John Moffat, Prof. Victor Brumberg,
Prof. Evgenij Edelman, Dr. Yurij Baryshev,
Dr. Andrzej Krasinski, Dr. Sergei Kopeikin, Dr.Sergei Krasnikov,
Dr.Roman Zapatrin and Marina Vasil'eva.
Dr. Krasinski sent me his book "Physics in an Inhomogeneous
Universe". \newline
This paper was financially supported by "COSMION" Ltd., Moscow.
\newline

{\bf REFERENCES}
\newline

Bondi H. (1947). MNRAS, {\bf 107}, 410.

Bonnor W.B. (1972). MNRAS, {\bf 159}, 261.

Bonnor W.B. (1974). MNRAS, {\bf 167}, 55.

Collins C.B. and Hawking S.W. (1973). AphJ, {\bf 180}, 317.

Gromov A. gr-qc/9612038

Gurevich L.E., Chernin A.D. (1978). The Introduction into Cosmology,

\qquad Moscow,"Mir".

Kopeikin S. 1996, private communication.

Ruban V.A., Chernin A.D. (1969). Proceeding of the 6th Winter

\qquad School on the Cosmophysics, p.15, Appatiti.

Silk J., (1977). Astronomy and Astrophysics, {\bf 59}, 53.

Tolman R.C. (1934). Proc.Nat.Acad.Sci (Wash), {\bf 20}, 169.

\end{document}